\documentclass[letterpaper,twocolumn,10pt]{article}
\usepackage[hyphens]{url}
\usepackage{usenix,epsfig,hyperref,caption,subcaption}
\hypersetup{
    bookmarksnumbered=true,
    unicode=false,          
    pdftoolbar=true,        
    pdfmenubar=true,        
    pdffitwindow=false,     
    pdfstartview={FitH},    
    pdftitle={Clome: The Practical Implications of a Cloud-based Smart Home},    
    pdfnewwindow=true,      
    colorlinks=false,       
    linkcolor=red,          
    citecolor=green,        
    filecolor=magenta,      
    urlcolor=cyan           
}

\begin{document}

\date{}

\title{\Large \bf Clome: The Practical Implications of a Cloud-based Smart Home}

\author{
{\rm Zubair Nabi}\\
IBM Research, Dublin\\
zubairn@ie.ibm.com
\and
{\rm Atif Alvi}\\
Forman Christian College, Lahore, Pakistan\\
atifalvi@fccollege.edu.pk
}

\maketitle

\begin{abstract}
A rich body of work in recent years has advocated the use of cloud technologies
within a home environment, but nothing has materialized so far in terms of
real-world implementations. In this paper, we argue that this is due to the fact
that none of these proposals have addressed some of the practical challenges of
moving home applications to the cloud. Specifically, we discuss the pragmatic
implications of moving to the cloud including, data and information security,
increase in network traffic, and fault tolerance. To elicit discussion, we take
a clean-slate approach and introduce a proof-of-concept smart home, dubbed
Clome\footnote{Short for CLoud hOME.}, that decouples non-trivial computation
from home applications and transfers it to the cloud. We also discuss how a
Clome-like smart home with decentralized processing and storage can be enabled
through OpenFlow programmable switches, home-centric programming platforms, and
thin-client computing.

\end{abstract}

\section{Introduction}
Today's home is a microcosm of personal computers, smart phones, embedded
devices, networked appliances, and sensors. All of these rely on highspeed
broadband and wireless connectivity. Highspeed Internet has also made rich
multimedia services such as Apple TV~\cite{AppleTV} possible. These living
spaces, which aim to improve quality of life are broadly categorized as
\emph{smart homes}. Smart homes invisibly perform tasks and functions that would
otherwise need to be performed manually by the users. 

With the widespread adoption of the vision of ``Internet of
Things''~\cite{Bohli:2009:IOE}, the number of connected devices is increasing
exponentially. We believe that the concept of ``Internet of Things'' blends well
with that of a smart home -- the home is the best place to start due to its
device diversity. The growing use of HTTP in the house~\cite{Maier:2009:DCR}
both by PCs and other embedded and mobile devices~\cite{Erman:2011:HHJ} is a
proof of this trend. All of this is in line with the vision of ubiquitous
computing but a number of outstanding issues need to be resolved before smart
homes can become truly pervasive. The existing home set-up is plagued by many
problems such as inflexibility and poor manageability of home automation
systems~\cite{Brush:2011:HomeAutomation}. Additional problems include high
energy consumption and heat dissipation, device heterogeneity, and poor
manageability. Adding more devices will only exacerbate these problems.

Similarly, home networks have their own set of problems. Provisioning,
complexity, troubleshooting, security, and composition problems can be
attributed to the use of the end-to-end principle in these
networks~\cite{Calvert:2007:movingtoward}. The absence of network administrators
in the home can also lead to \emph{misconfiguration} of the
network~\cite{Aggarwal:2009:NDH}. Further, home networks are also susceptible to
spam, \emph{DoS} traffic, and/or scam or phishing
attacks~\cite{Feamster:2010:OHN}. A number of solutions to these problems have
been
proposed~\cite{Calvert:2007:movingtoward,Aggarwal:2009:NDH,Feamster:2010:OHN}
but each one of them requires dedicated resources and intensive computation in
some cases.

Concurrently, researchers have started thinking of the home as a platform with
its own Appstore to serve applications ranging from security systems to light
controllers~\cite{Dixon:2010:HNO}. In the same vein, software defined networking
(SDN) has also opened up a market for networking applications ranging from QoS
to identity management~\cite{Merchant:2012:LAS}. But there is still no common
platform and application store to house all of these efforts under one roof.

The emergence of cloud computing has revolutionized large-scale computation,
storage, application development, and web services. Enterprises and service
providers no longer need to spend directly on hardware, software, maintenance,
and manpower. They can outsource them to cloud vendors and be functional in a
matter of minutes with fine-grained control over scalability.
The key advantages include on-demand scaling, pay-as-you-go price model, and
high-speed network access~\cite{Li:2010:CSC,Popa:2010:CTA}.

In view of the discussion above, it seems natural that a number researchers have
advocated the use of the cloud to implement smart
homes~\cite{Wei:2010:DCA,Yang:2010:ACA,Ye:2011:AFA,IBM:2010}. But not
surprisingly, none of these proposals have been able to garner much traction. We
argue that while the advantages of migration to a cloud have been highlighted
\emph{ad infinitum}, the challenges have largely been ignored. We believe that
these challenges and practical considerations need to be addressed before a
cloud-driven smart home can become a reality. These challenges range from an
increase in network traffic and opaque security to weak fault-tolerance and
non-trivial cost model. Therefore, the major contribution of this paper is a
discussion of these problems and where possible, a proposition of potential
solutions. To drive the discussion, we also present the design and architecture
of a cloud-aware smart home, dubbed \emph{Clome}. The design of Clome is general
enough to be applicable to any smart-home setup and realistic enough to be
readily implemented and deployed.

\textbf{Structure of the Paper:}
We present the motivation behind cloud-based smart homes in
\S\ref{sec:motivation}.\S\ref{sec:clome} presents the architecture of Clome and
proposed implementation details. Applications that can benefit from a cloud
based implementation are enumerated in \S\ref{sec:applications}.
\S\ref{sec:challenges} highlights the challenges that arise due to a cloud
environment. \S\ref{sec:relatedWork} gives an outline of related work. We
conclude in \S\ref{sec:conclusion} and also discuss future directions.

\section{Motivation}\label{sec:motivation}
To motivate the discussion, this section lists a number of advantages of moving
home applications to the cloud.

\textbf{Application Development:} 
Development on the cloud will shorten development and prototyping
time~\cite{IBM:2010}. Also, the use of a common execution platform will lead to
common programming practices and design standardization.

\textbf{Device Heterogeneity:} 
At present, homes contain a large number of diverse devices which complicates
application development and
interoperability~\cite{Dixon:2010:HNO,Brush:2011:HomeAutomation}. The use of
thin-client computing where all the computation is performed on the cloud and
only the output is displayed in the home, will lead to device homogeneity.

\textbf{Flexibility and Scalability:}
Integrated smart home systems are inflexible, while on the other hand using
multiple vendors can lead to interoperability
issues~\cite{Brush:2011:HomeAutomation}. Thus, changing or adding a new device
requires a number of strict design choices. Using \emph{dumb} client-side devices
in the home makes the entire system flexible. Additionally, software upgrades can
be applied to applications without a need to update the firmware in the
thin-client. Further, the requirements of new applications can be fulfilled by
scaling up the cloud resources.

\textbf{Energy Efficiency:} In 2005, households accounted for 42\% of energy
consumption in the European Union~\cite{Tompros:2008:PNA}. Electricity
consumption can be reduced as data centers effectively use smart load
management~\cite{Parolini:2008:RDC}, efficient
architecture~\cite{Gyarmati:2010:AHR}, and energy-aware
routing~\cite{Shang:2010:ERD}. Moving computation to the cloud also moves heat
dissipation to the cloud where data centers make use of techniques such as
high-efficiency water-based cooling systems to act as heat
sinks~\cite{Katz:2009:Tech}.

\textbf{Pricing:} Users can also take advantage of both temporal and geographic
fluctuation in electricity prices to reduce their bill~\cite{Qureshi:2009:CEB}.
Studies have shown that government agencies that have moved to the cloud have
been able to make savings between 25 to 50 percent~\cite{West:2010}. Similarly,
moving compute intensive applications from the home to the cloud will also lead
to comparable gains.

\textbf{Mobility:} As more users adopt this model, standardized devices will
become a part of every home. Users can move residence and start using their
existing applications as their account/profile would be stored on the cloud.

\textbf{Common User Interface:} Current user interfaces in the home are too
complicated for the average user. The use of a common user interface through
which users can interact with the system and give their input to devices makes
the environment accessible to everyone~\cite{Calvert:2007:movingtoward}. In
addition, the current desktop-centric interfaces are incompatible with
cloud-backed applications where people take center stage~\cite{Pham:2011:UIM}.
Decoupling the computation infrastructure (by moving it to the cloud) and the
input system, enables multiple user interfaces to exist side by side allowing
user-centric customization.

\textbf{Network Management and Security:} The use of a common processing
platform allows users to take advantage of applications that can ensure proper
network configuration~\cite{Aggarwal:2009:NDH}, reduce network
complexity~\cite{Calvert:2007:movingtoward}, and provide security against
attacks and spam~\cite{Feamster:2010:OHN}. The synergy of these applications
will allow users to get the most out of their networks.

\section{Clome}\label{sec:clome}
This section presents the architecture of Clome followed by a proposed
implementation.

 \begin{figure*}[t]
  \centering
    \includegraphics[width=4.5in]{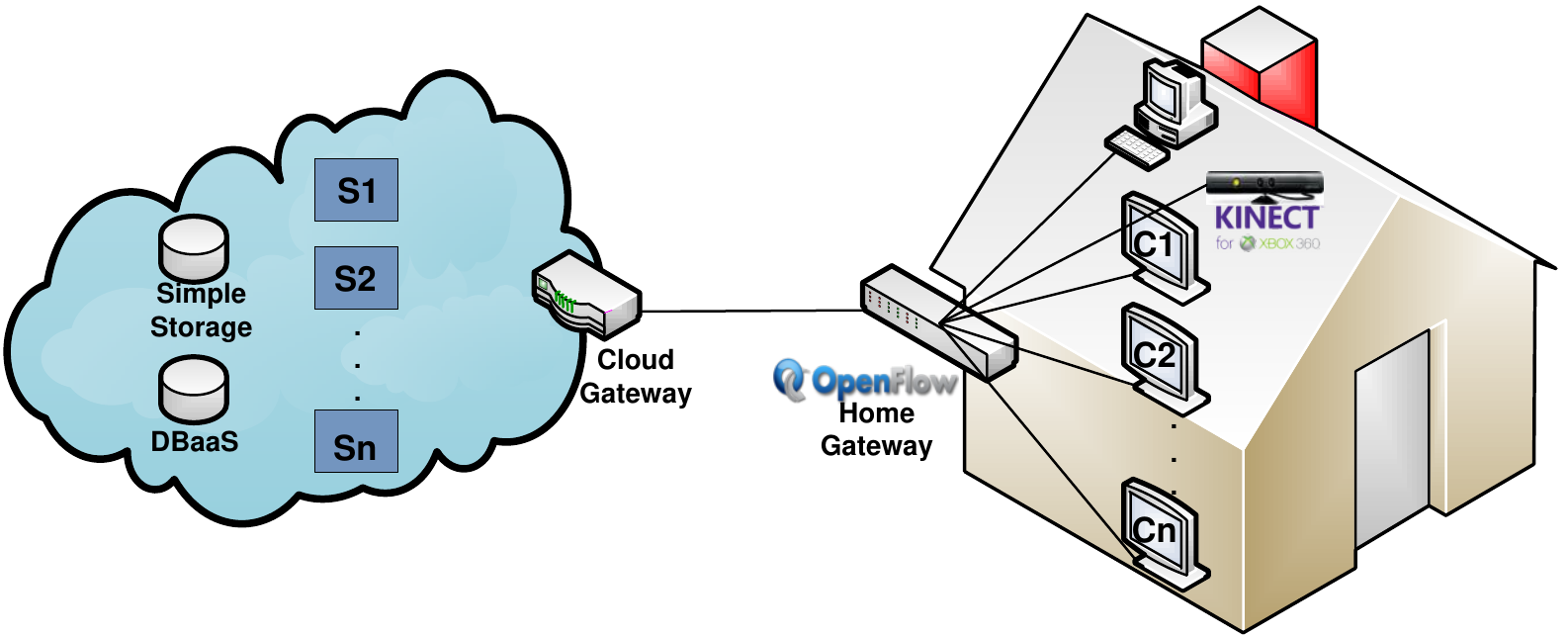}
  \vspace{-10pt}
  \caption{Architectural overview of Clome. S1 to Sn
  and C1 to Cn represent the server and client ends of
  thin-client applications, respectively.}\label{fig:architecture}  
\end{figure*}
\subsection{Architecture}
We present one possible\footnote{We do not claim that this is the best
architecture for a cloud-enabled smart house but just a good starting point to
elicit discussion.} architecture in Figure~\ref{fig:architecture}. All entities
inside the house are connected to the outside world through a programmable
network switch. All applications and smart appliances have a CPU-heavy end
staged in the cloud represented by \emph{S} and a thin-client end represented by
\emph{C}. Applications in the cloud can make use of both simple storage and a
transactional database. Users interact with and control the system and all
applications through a natural user interface. Traditional devices such as PCs,
smart phones, and tablets are also connected to the same network and can also be
used to access applications staged on the cloud.

\subsection{Proposed Implementation}
The gateway is enabled by an OpenFlow compliant switch~\cite{McKeown:2008:OEI}
connected to a high-speed Internet backbone. For thin-client computing, we
advocate the use of THINC~\cite{Baratto:2005:TVD}, a high-performance system
that virtualizes the display at the device driver level. It uses a small set of
low-level display commands that reduce both computation and bandwidth overheads.
In addition, it implements a number of optimizations to ensure high
interactivity even in low-latency networks. The natural user interface in this
architecture is enabled by Microsoft's Xbox Kinect~\cite{Kinect}. It has a
built-in RGB camera, depth sensor and multi-array microphone. These enable
Kinect to get user input through visual gestures and voice commands.

\section{Applications}\label{sec:applications}
In this section we describe both new applications and existing ones from the
literature which can benefit from a cloud based implementation. The applications
can be made available to users through a \emph{Home
Store}~\cite{Dixon:2010:HNO}. These applications can be categorized into 2 broad
categories: 1) User-centric, and 2) System-level. We discuss the former first
followed by the latter.

\subsection{User-centric Applications}

\subsubsection{Storage}\label{storage}
Home users need multiple media such as compact discs and specialized storage
drives to store their music collections. They also buy sophisticated set-top
boxes and media centers to store their favourite movies and TV shows. This is
both expensive in terms of hardware and extremely inflexible, i.e. there's an
upper bound on the storage capability of a device. On the other hand, storing
all this media in the cloud is both cost effective and scalable. Applications
can directly stream favourite media from the cloud. Transactional databases can
be used to implement applications that require structured data such as
phonebooks and large music catalogues.

\subsubsection{Gaming}
According to a study~\cite{Softpedia}, two-thirds of US households own a gaming
console. Out of these, 65\% of houses also have broadband connectivity.
Additionally, it is expected that 260 million consoles will be connected to the
Internet by 2015~\cite{Examiner}. These consoles have high processing speed,
memory and local storage. For example, the Xbox 360 has 3 CPU cores which run at
3.2 Ghz, 512 MB DRAM, a 500 Mhz GPU and optional storage of up to
500 GB~\cite{Andrews:2006:XSA}. A cloud version of these consoles would perform
its processing on the cloud and its output would be streamed directly to
television sets. This \emph{cloud console} would receive its input over the
network from the Kinect controller. Multi-player capability is enabled through
the use of Virtual LANs in the cloud. Such an approach would make the purchasing
of new upgraded consoles redundant as the cloud resources can easily be scaled up
with the emergence of newer games.

It is also important to highlight that a cloud-based gaming model greatly
simplifies the architecture of Massively Multiplayer Online Games (MMOGs). Such
cloud-based systems can gracefully service millions of concurrent
users~\cite{Marzolla:2011:NGGI}.

\subsubsection{Automated Surveillance and Voice Recognition}
Surveillance systems are one of the most popular components of a smart home. They
can be integrated with the security system to provide new functions such as
automatically opening the door when a home owner is at the door using face
recognition. Additionally, these systems can also be voice activated and can
discern the subtleties of natural human language to perform different tasks.
Unfortunately, computer vision and natural language processing algorithms are
compute intensive. The solution lies in the model of cloud enabled robotics which
offload CPU-intensive tasks including speech recognition and 3D mapping to the
cloud~\cite{Arumugam:2010:DAvinCi}. Such a system can also make use of
distributed data intensive computing systems on the cloud for face recognition
and other computer vision tasks~\cite{White:2010:WCV}.

\subsubsection{Health Monitoring}
Services such as Vignet~\cite{Vignet} provide personalized health care solutions.
Such services can be made a part of every home and their functionality can be
enhanced by using the cloud. New health applications can use data from multiple
sensors located in the home. Applications in the cloud can then mine the data to
look for unusual patterns. These applications raise a flag when something
abnormal happens to alert paramedic services. Privacy is ensured as the data is
in the cloud and third-parties only get alerts.

\subsection{System-level Applications}

\subsubsection{Security}
Poor management and lack of technical know-how make home networks easy targets
for a wide range of attacks. Also, systems in these networks can become
compromised and become a source of spam and malicious
traffic~\cite{Feamster:2010:OHN}. One suggested approach~\cite{Feamster:2010:OHN}
to tackle this problem involves the use of OpenFlow programmable networking
switches~\cite{McKeown:2008:OEI} and distributed inference algorithms in a closed
feedback loop. The OpenFlow switch is used to both forward traffic to the
inference engine and to implement new traffic filtering rules. These rules are
computed by the distributed inference engine that uses various spam filtering and
botnet/malware detection algorithms. In case of Clome, this third party
inference engine -- because of its compute-intensive nature -- is deployed in the cloud
from where it can control the OpenFlow gateway switch. Users of this application
form region-wise hierarchical Virtual LANs to be a part of the distributed
inference.

\subsubsection{Misconfiguration}
 As mentioned earlier, the home network lacks a local administrator who can
 properly configure the large number of diverse applications and devices within
 it. \emph{Misconfiguration} is a problem which arises due to the inexperience of
 the home user or the interaction of various components with each
 other~\cite{Aggarwal:2009:NDH}. Such misconfiguration can easily bring down the
 home network. NetPrints~\cite{Aggarwal:2009:NDH} is a tool which leverages the
 fact that there must be at least one user in the network with a correct
 configuration, and uses that information to properly configure misbehaving
 elements. It automatically indexes and retrieves this shared knowledge. Users
 label configurations as either ``good'' or ``bad''. NetPrints has a
 client/server model, in which the client runs on all elements in the network and
 collects configuration information and a network traffic trace. It uploads this
 information to a remote server which uses tree based learning and a
 \emph{mutation algorithm} to construct a tree of both working and faulty
 configuration settings and gives suggested fixes to the clients. As the number
 of clients increases, the server can become a performance bottleneck. Therefore,
 Clome hosts the server on the cloud.

\subsubsection{Traffic Monitoring}
The popularity of cloud-managed routers from companies such as
Meraki~\cite{Meraki} has underlined the power of decentralized management and
control. These routers enable features such as per-application QoS and traffic
analysis. But they are designed for large enterprises with up to 5000 sites
which makes them both unfeasible and unaffordable for home users. Interestingly,
studies have shown that making per-user usage information available to all
households members can help in better management of
bandwidth~\cite{Chetty:2010:WHB}. Additionally, decentralized monitoring can
automatically limit the bandwidth of both users and applications. Such a system
can also be used to implement parental controls, in which undesirable content
can be blocked. Therefore, Clome uses commodity OpenFlow
switches~\cite{McKeown:2008:OEI} and monitoring algorithms running on the cloud
to enable traffic monitoring and management.

\subsubsection{Smarter Network}\label{smarterNetwork}
Problems that the previous three applications address can also be attributed to
the inherent mismatch between the home environment and the architecture of the
Internet~\cite{Calvert:2007:movingtoward}. The reliance of home networks on the
end-to-end principle leads to end-host problems. Calvert et
al.~\cite{Calvert:2007:movingtoward} propose a ``smart middle'' approach which
turns the home network into an ``edge network''. This approach simplifies packet
forwarding, device monitoring and brokering, and policy management. It makes use
of a central \emph{portal} as the control entity which provides connectivity,
policy management, and mediation between devices. The portal controls both intra
and inter home communication through a managed switch called the
``interconnect''. The portal also maintains a database of device location and
behaviour. In Clome the portal is housed in the cloud and an OpenFlow
switch~\cite{McKeown:2008:OEI} is used as the interconnect.

\section{Practical Implications}\label{sec:challenges}
In this section we highlight some challenges that arise as a result of moving
home applications to the cloud. We also suggest possible solutions to these
challenges.

\subsection{Cost Model}\label{sec:cost}
Assuming a \emph{Large Instance}\footnote{7.5 GB memory, 4 EC2 Compute Units,
850 GB instance storage, 64-bit platform.} from Amazon EC2, a quick
back-of-the-envelope calculation reveals that such an instance would cost
approximately \$245 per month at the current per hour
rate\footnote{\url{http://aws.amazon.com/ec2/pricing/}}. Moving roughly 1 TB of
data in and out of the cloud per month would incur an additional cost of \$100.
Further, 1 TB of permanent storage in the Elastic Block Store adds another
\$100. Factoring in I/O request costs and location variation, the total monthly
cost comes out to be roughly \$500. This is a reasonable price to pay for the
large number of benefits highlighted in \S\ref{sec:motivation}. Additionally, we
believe that as cloud usage by home users increases, cloud vendors will
introduce monthly package deals, in the same way as Internet charges have
evolved from per hour pricing to a monthly flat-rate.

\subsection{Choosing a Cloud Vendor}
In a home environment, tenants have the luxury of choosing vendors for different
services such as cable TV or broadband access. But in contrast, choosing a cloud
vendor is a non-trivial task due to: 1) The different charging models of cloud
services. For example, some charge by the hour while others by computation
cycles, 2) The computation behaviour of different applications. For example, some
applications might be compute intensive while others might be I/O bound.
Fortunately, CloudCmp~\cite{Li:2010:CSC} is a framework which can assist users in
directly choosing a platform. The same framework can also be used by application
developers to make recommendations to users.

\subsection{Cloud Vendor API Lock-in}
A key selling point of cloud-enabled smart homes is mobility: tenants can change
residence and their applications and profiles follow suit. This vision is in
tension with cloud vendor API lock-in wherein applications are designed around
specific APIs and any change in the API requires a complete application
re-design~\cite{Armbrust:2010:VCC}.  To abstract away the API and low-level
details of each cloud vendor, we propose the use of abstraction libraries such
as Apache LibCloud~\cite{LibCloud}, an opensource client library to standardize
the interaction of applications with cloud services.

\subsection{Copyright}
The cloud has been touted as a convenient storage container for personal audio,
image, and video collections. At the same time, the intellectual property rights
side of the picture has not received the required attention. For instance, most
image collections consist of pictures taken by the users themselves. In
contrast, audio and video collections largely contain music, movies, and TV
shows either purchased online or ripped from CDs, DVDs, etc. In some cases, some
of this content is illegally downloaded from P2P file sharing networks. Services
such as Apple's iCloud deal with illegal music by virtually legalising it for a
small yearly premium~\cite{iCloud:plag}. We believe that the situation will be
complicated by the addition of a diverse of content. Legal experts will need to
be brought on-board before such storage containers see the light of day.

\subsection{Inter-Application Communication}
 In some cases it might be necessary for two applications to communicate with
 each other. For example, a network management system might need regular traffic
 information from the security application. But the situation is complicated if
 these applications reside inside different Virtual Machines.
 CloudPolice~\cite{Popa:2010:CTA}, an end-host hypervisor access control system
 can be used to implement privacy and security. In addition, \emph{feature
 interaction} between different applications might result due to
 inter-application communication. Feature interaction is beyond the scope of this
 paper but the approach described in~\cite{Greaves:2008:U.C} using bytecode
 profiles can easily be ported to a cloud environment.
 
\subsection{Increase in Home Network Traffic}
Communication between the client and server ends of applications in Clome
imposes an additional network traffic overhead. But with reported broadband
speeds of up to 17/1.2 Mbps (downlink/uplink) in some
residences~\cite{Maier:2009:DCR}, we believe that current and obviously future
networks will easily be able to support such traffic. In addition, different
compression and encoding techniques can be used in tandem as well. Furthermore,
a local caching server can also be deployed inside the home which can cache
content which is popular across devices. Another challenge is network allocation
and QoS for different applications. Standard network traffic is predominantly
TCP and UDP based but streaming applications implement their own custom
protocols such as RTP. Realistically, different applications will use different
transport and application level protocols to best suit their requirements and
the interaction of all them on the same network link is an open problem.

\subsection{Security}
Providing security is an overarching goal as a security breach in a home control
system can be fatal. One can easily imagine an adversary gaining control over
applications or accessing private data. This can happen both on the wire and at
the cloud end. In case of the former, lightweight encryption at any level in the
network stack (IPsec (L3), tcpcrypt (L4), or TLS/SSL (L7)) can be used to
protect information. The situation is more complicated at the cloud end as data
security in the cloud is a challenging problem but at the same time the cloud is
uniquely positioned to enable confidentiality and
auditability~\cite{Armbrust:2010:VCC}. So much so that even healthcare companies
with sensitive patient records have already moved to the cloud. In addition,
CryptDB-like~\cite{Popa:2011:CPC} systems enable queries to be run over
encrypted data so that even cloud providers and system administrators cannot
access it to ensure privacy. Finally, virtual machines can sandbox the user's
data and computation within a single domain and make them inaccessible to
malicious users.

\subsection{Fault Tolerance}
Datacenters which house the cloud are failure-prone due to a number of reasons,
including component and power failure. Very recently Amazon Web Services (AWS)
experienced problems leading to a blackout of several major websites and
services~\cite{AWS:2012:Failure}. This can be disastrous for a smart home. A
number of solutions to this problem can be employed. First and foremost, cloud
services are geo-replicated so that if one site fails, computation and storage
can continue from another. Secondly, users have the option of subscribing to a
secondary (backup) cloud vendor, which adds another line of defense against
failure. Finally, there can be a dichotomy between critical and non-critical
applications, with the former provided failure redundancy within the home
environment itself (more details later).

Another problem that can plague a cloud-enabled smart home is Internet failure.
In such cases, a backup network subscription can also be maintained. Moreover,
in the common case, each home would be represented by a single VM staged in the
cloud. As most applications are susceptible to software bugs, the entire VM can
crash due to a single application. To avoid such a situation, applications can
be housed within lightweight OS-level containers, to ensure failure and
performance isolation.

\subsection{Opaque Cloud Setup}
While subscribing to a secondary cloud vendor is one way to ensure
fault-tolerance, this too is not foolproof as cloud vendors at times depend on
services provided by the same third party~\cite{Ford:2012:ICO}. For instance,
two cloud vendors might be dependent on storage provided by the same storage
provider. Therefore, their failure will be correlated. Unfortunately, there is
no mechanism for the application to have such insight as these details are
abstracted away by the cloud stack. One possible solution for this problem is
for each vendor to expose an explicit dependency graph which can be used by the
rich set of smart-home applications to reason about the underlying
structure~\cite{Ford:2012:ICO}, thus negating this failure correlation.

\subsection{Everything in the Cloud?}\label{subsection:everything}
So far we have assumed that all applications can be moved to the cloud with
potential benefits. At the same time, it is also important to acknowledge that
moving some trivial applications to the cloud might be overkill. For instance,
moving the logic of a light dimmer switch seems highly unnecessary. Therefore,
we are working on defining a generic metric which users and developers can
employ to find out if an application can benefit from a cloud-based
implementation. For critical applications and those with no visible benefits
from a migration to the cloud, we envision a centralized HomeOS~\cite{Dixon:2010:HNO}
server\footnote{This server need not be a full-fledged machine. For most
simplistic applications, a Raspberry Pi~\cite{RaspberryPi} will suffice.}
running within the house. Additionally, this server can also be used to access
and configure the cloud instance. Furthermore, to ensure complete privacy, users
can store their confidential data locally and grant access to remote
applications on a need to know basis. Finally, the same local server can receive
input from data-intensive sensors and only forward a periodic snapshot to the
cloud for any non-trivial processing.

\section{Related Work}\label{sec:relatedWork}
In this section we present relevant related work which inspired us to write this
paper.

HomeOS~\cite{Dixon:2010:HNO} is a home-centric operating system designed to
negate device heterogeneity and to provide a common application development
platform. Further, these applications can be downloaded from a \emph{HomeStore}
on the same lines as the Apple \emph{App Store}. In the same vein, Clome
applications can easily be downloaded from a store to the user's cloud
resources. In addition, as mentioned in \S\ref{subsection:everything}, a local
HomeOS can be used to run applications for which a cloud migration would be
overkill. Niedermayer et al.~\cite{Niedermayer:2009:UHN} have tried to devise
strategies for the \emph{Future Internet}. Out of these, the strategies most
relevant to our work are, a) defining a well known boundary for a home network
and all the devices and users within it and, b) providing centralized network
resources that make use of cloud computing and P2P networks for data storage and
context-dependent access. Clome is complementary to their work as it deals with
the \emph{Future Home} by making use of the same design principles.

A cloud based smart home is also described by
~\cite{Wei:2010:DCA,Yang:2010:ACA,Ye:2011:AFA}. Unfortunately none of them
discuss the benefits of such a set for a home environment or any security
implications. Similarly, Reference~\cite{IBM:2010} envisions a \emph{smarter}
home that makes use of the cloud for faster service development and deployment
cycle but does not address any practical deployment issues.

\section{Conclusion and Future Work}\label{sec:conclusion}
Using a proof-of-concept cloud-enabled smart home, dubbed Clome, we argued that
a number of practical issues need to be addressed before such architectures can
find real-world deployment. These issues include cloud vendor API lock-in,
security, fault-tolerance, and network traffic. We took the position that some
of these issues can be easily tackled using existing solutions but others
require attention from the research community. We also highlighted some of the
advantages of moving applications to the cloud, such as minimizing the
computation and energy footprint. Furthermore, our design of Clome builds upon
existing technologies, thus enabling it to be readily implemented and deployed.

As Clome is a work-in-progress, our future work is extensive. It includes the
testing and evaluation of all applications described in this paper. We also
intend on deploying these applications in actual homes to study the effect of
the bandwidth overhead. Further, we are also interested in designing thin-client
components that can seamlessly be embedded in existing appliances.

{\footnotesize \bibliographystyle{acm}
\bibliography{clome}}

\end{document}